\documentclass[onecollarge]{svjour2}       % onecolumn "king-size"
\smartqed  % flush right qed marks, e.g. at end of proof
\usepackage{graphicx}
\usepackage{bm}
\usepackage{amssymb,amsmath}
\usepackage[usenames]{color}
\usepackage{epsfig}
\usepackage{hyperref}
\usepackage{subfigure}
\usepackage[sans]{dsfont}
\hypersetup{colorlinks=true, citecolor=blue,
linkcolor=blue,urlcolor=blue }
\begin{document}

\title{Ground state properties of Ising chain with random monomer-dimer couplings
}


\author{S. Bahareh Seyedein Ardebili         \and
        Reza Sepehrinia
}

\authorrunning{S. B. Seyedein Ardebili \and R. Sepehrinia} % if too long for running head

\institute{\email{sepehrinia@ut.ac.ir}\\ \\
S. Bahareh Seyedein Ardebili \at
              Department of Theoretical Physics and Astrophysics, University of Tabriz, Tabriz 51666-16471, Iran \\
%              \email{fauthor@example.com}           %  \\
           \and
            Reza Sepehrinia \at
              Department of Physics, University of Tehran, Tehran 14395-547, Iran,  and \\
              School of Physics, Institute for Research in Fundamental Sciences, IPM, 19395-5531 Tehran, Iran
}


\maketitle

\begin{abstract}
We study analytically the one-dimensional Ising model with a random binary distribution of ferromagnetic and antiferromagnetic exchange couplings at zero temperature. We introduce correlations in the disorder by assigning a dimer of one type of coupling with probability $x$, and a monomer of the other type with probability $1-x$. We find that the magnetization behaves differently from the original binary model. In particular, depending on which type of coupling comes in dimers, magnetization jumps vanish at a certain set of critical fields. We explain the results based on the structure of ground state spin configuration.
%Include keywords, PACS and mathematical
%subject classification numbers as needed.
\end{abstract}

\section{Introduction}
\label{intro}
Ground state magnetization of random-bond Ising chain, described by the Hamiltonian,
%$\mathcal{H}=-\sum_{i} J_i \sigma_i \sigma_{i+1}-h\sum_i \sigma_i$ with $J_i=\pm J$,
%
\begin{equation}\label{Ising}
\mathcal{H}=-\sum_{i} J_i \sigma_i \sigma_{i+1}-h\sum_i \sigma_i,
\end{equation}
with the binary random distribution of $J_i=\pm J$,
exhibits a staircase behavior as a function of the magnetic field with a sequence of steps at $h=2J/r$ with $r=1,2,\cdots$, and flat plateaus in between \cite{derrida1978simple,vilenkin1978random}. Transitions between adjacent plateaus are accompanied by a drastic change in the spin configuration. Entropy has a similar behavior
except at the transitions, where it shows sharp spikes indicating a highly degenerate state at these points. This unusual behavior allows, however, a simple explanation. At the critical fields the energy cost, $2J$, for flipping clusters of down spins of a certain length, $r$, is equal to the energy gain, $-2rh$, from the magnetic field. This results in many degenerate configurations. At the slightly higher magnetic field, flipping of such clusters becomes favorable and a jump occurs in the magnetization. At very small values of magnetic field, magnetization undergoes an infinite number of discontinuities, but varies linearly with the magnetic field and the susceptibility has a finite limit. This property will be modified if the $J_i$s are taken from a continuous distribution. For a continuous distribution, including neighborhood of $J=0$, it is shown \cite{chen1982low,gardner1985zero} that magnetization has nonanalytic scaling behavior in weak magnetic field, $M\sim h^{\mu}$, where $\mu=\frac{y+1}{y+3}$ is a non-universal exponent depending on the shape of distribution near the zero coupling $\rho(J\rightarrow0)\sim |J|^{y}$. In particular with a Gaussian distribution ($y=0$), magnetization behaves as $M\sim h^{\frac{1}{3}}$. In contrast with the binary distribution, magnetic susceptibility diverges as $h\rightarrow 0$ with this type of distributions.

Besides the distribution function of couplings, it is known that correlations in the distribution can also markedly change the macroscopic properties of the Ising model \cite{nieuwenhuizen1987exact,ballesteros1999site,bagamery2005two,rajabpour2008explicit} and the collective behavior of statistical models, in general, \cite{weinrib1983critical,weinrib1984long,prudnikov2000field}. In this paper, we study the binary model with a correlated distribution of couplings. We introduce correlations in the disorder by assigning a dimer of one type of coupling with probability $x$, and a monomer of the other type with probability $1-x$. The correlation length of the couplings depends on the concentration of dimers which allows us to have an adjustable range of correlation. We find analytical results for ground state energy and magnetization and discuss them in comparison with the original model with a uncorrelated distribution of couplings.

\section{Method and macroscopic quantities}

As we mentioned above, the behavior of magnetization can be explained by studying the clusters of down spins that flip as the magnetic field increases. However, one needs to know the number of such clusters to calculate the magnetization change, entropy, and other macroscopic properties. We only mentioned flipping a single cluster, but there are other possibilities. For example, two clusters of down spins which are separated by a cluster of up spins can turn up while the up spins in between turn down in some critical field. In other words, a cluster of given length could flip in more than one critical field. This makes the problem of counting the clusters, that flip in a given critical magnetic field, more complicated. However, the analytical calculation of magnetization and the entropy can be performed without actually counting the clusters. This has been done by using the transfer matrix approach \cite{derrida1978simple} and by mapping the problem onto a random walk \cite{vilenkin1978random}. Here we use the former approach, which is described in detail in \cite{derrida1978simple,crisanti1993products}, and briefly, introduce the formalism that we will be using throughout this paper. For a chain of $N$ spins, the partition function can be expressed by a matrix product
\begin{eqnarray}
  Z = \sum_{\{\sigma_i\}} e^{-\beta \mathcal{H}} = \text{Tr} (\mathcal{L}_1 \mathcal{L}_2 \cdots \mathcal{L}_N)\equiv \text{Tr} {\mathcal{P}_N}, \label{partition}
\end{eqnarray}
where $\mathcal{L}_i$s are ${2}\times{2}$ matrices with elements $\mathcal{L}_i(\sigma_i,\sigma_{i+1})=\exp(\beta J_i \sigma_i \sigma_{i+1}+h\sigma_i)$. With uncorrelated couplings $J_i$, which are taken with the probability distribution $\rho(J_i)=(1-x)\delta(J_i-J)+x\delta(J_i+J)$, we have two kinds of matrices:
\begin{eqnarray}
  \mathcal{L}_i =\left\{\begin{array}{l}
         \left(\begin{array}{cc}
            z^{1+\alpha} & z^{-1+\alpha} \\
            z^{-1-\alpha} & z^{1-\alpha}
         \end{array}\right)  \text{with prob.} \ 1-x  \vspace{0.1 cm}\\
         \left(\begin{array}{cc}
            z^{-1+\alpha} & z^{1+\alpha} \\
            z^{1-\alpha} & z^{-1-\alpha}
          \end{array}\right)  \text{with prob.} \ x \
             \end{array}  \right.
\end{eqnarray}
where $z=\exp(\beta J)$, $\alpha=h/J$ and the matrices correspond to ferromagnetic ($J_i=J$) and anti-ferromagnetic ($J_i=-J$) couplings respectively. We will call this model AF throughout this paper where the name indicates the types of couplings which are distributed along the chain, for this case anti-ferromagnetic (A) and ferromagnetic (F) monomers. Free energy per spin, in the thermodynamic limit, is given by
\begin{equation}
F=-\frac{1}{N\beta }\ln Z=-\frac{1}{\beta}\left\langle\lim_{N\rightarrow\infty} \frac{1}{N} \ln \text{Tr} {\mathcal{P}_N}\right\rangle,
\end{equation}
where the average is taken over all the possible configurations of couplings. The right-hand side is indeed proportional to the Lyapunov exponent of the matrix product. Since the matrices $\mathcal{L}_i$ have positive elements, it can be verified that the trace can be replaced with any element of $\mathcal{P}_N$. At the low-temperature limit, i.e., $z\rightarrow\infty$ it is sufficient to keep for each element of $\mathcal{P}_N$ the leading term in $z$ which in general would be of the form $z^{a}e^{u}$. Therefore, at this limit, the free energy is given by
\begin{eqnarray}
F\simeq-\frac{1}{\beta}\lim_{N\rightarrow\infty}\left(\frac{\langle a\rangle_{N}}{N}\ln z+\frac{\langle u\rangle_{N}}{N}\right),
\end{eqnarray}
and using $z=e^{\beta J}$, $\beta=k_BT$ the ground-state energy and entropy per spin read as
\begin{eqnarray}
E&=&-\frac{1}{N}\frac{\partial \ln Z}{\partial \beta}=J \lim_{N\rightarrow\infty}\frac{\langle a\rangle_{N}}{N}=J \lim_{N\rightarrow\infty}(\langle a\rangle_{N+1}-\langle a\rangle_{N}), \label{energy}\\
S&=&-\frac{\partial F}{\partial T}=k_B\lim_{N\rightarrow\infty}\frac{\langle u\rangle_{N}}{N}
=k_B\lim_{N\rightarrow\infty}(\langle u\rangle_{N+1}-\langle u\rangle_{N}). \label{entropy}
\end{eqnarray}
The last equality in Eq. (\ref{energy}), and similarly in (\ref{entropy}), follows from the identity $\lim_{N\rightarrow\infty}\frac{\langle a\rangle_{N+1}}{N+1}=\lim_{N\rightarrow\infty}\frac{\langle a\rangle_{N}}{N}$, and can be reexpressed as a single average $\langle a_{N+1}-a_N\rangle=\langle \Delta a \rangle$ where $\Delta$ is the first order difference operator such that $\Delta a=a_{N+1}-a_N$.

To calculate the averages in  (\ref{energy}), (\ref{entropy}) one needs the probability densities of the exponents $a,u$. A system of coupled stochastic recursive equations for the elements in each row of $\mathcal{P}_N$ can be obtained using
\begin{equation}\label{rec-eq}
\mathcal{P}_{N+1}=\mathcal{P}_N \mathcal{L}_{N+1}.
\end{equation}
From this equation, the elements in the first row of $\mathcal{P}_{N+1}$ are related to the elements in the first row of $\mathcal{P}_N$ so we only need to keep the first row in the iteration
\begin{eqnarray}
 \mathcal{P}_N \simeq \left(\begin{array}{cc}
            z^{a}e^{u} & z^{b}e^{v} \\
            ... & ...
          \end{array}\right), \hspace{0.5cm}
 \mathcal{P}_{N+1} \simeq \left(\begin{array}{cc}
            z^{a'}e^{u'} & z^{b'}e^{v'} \\
            ... & ...
          \end{array}\right).
\end{eqnarray}
For the exponents we have the joint probability density $\rho_{N}(a,b,u,v)$. The corresponding exponents of $\mathcal{P}_{N+1}$, i.e., $a',b',u',v'$ can be obtained in terms of $a,b,u,v$ for every choice of $\mathcal{L}_{N+1}$. We will see that knowing $\mathcal{L}_{N+1}$ and $c=a-b$ is enough to determine $a',b'$ and further with the knowledge of $w=u-v$, one can determine $u',v'$. Therefore, we may express the averages in terms of reduced probability densities $p_N(c)$, $p_N(c,w)$
\begin{eqnarray}
\langle \Delta a\rangle &=& \langle a'- a\rangle =\sum_{c,\xi} f_{\xi}(c) p_N(c), \label{ava} \\
\langle \Delta u\rangle &=& \langle u'- u\rangle =\sum_{c,w,\xi} g_{\xi}(c,w) p_N(c,w), \label{avu}
\end{eqnarray}
where, $p_{N}(c,w)=\int\rho_{N}(a,b,u,v)\delta(a-b-c)\delta(u-v-w) da \hspace{0.03cm} db \hspace{0.03cm} du \hspace{0.03cm} dv$, $p_N(c)=\int p_N(c,w) dw$. Functions $f_{\xi}$ and $g_{\xi}$ represent $\Delta a$ and $\Delta u$ respectively as a function of $c$ and $w$ upon multiplication of transfer matrices associated with $\xi=x$ or $\xi=1-x$. We will also see that $c$ and $w$ take specific values and when $N\rightarrow\infty$, the corresponding probability densities, tend to limit densities.
\section{Monomer-dimer chain}
Here we consider two different distributions of, $\pm J$, couplings. One is the distribution of ferromagnetic dimers and antiferromagnetic monomers and the second is the distribution of antiferromagnetic dimers and ferromagnetic monomers. We will call these cases AFF  and AAF, respectively, throughout the paper. The same method that we described above for AF model can be applied to the present cases by replacing the corresponding transfer matrix of repeated link with its squared matrix
\begin{eqnarray}
 \text{AFF:} \hspace{0.5 cm} \mathcal{L}_i &=& \left\{\begin{array}{l}
               \left(\begin{array}{cc}
            z^{2+{2}\alpha}+z^{-2} & z^{{2}\alpha}+1 \\
            1+z^{{-2}\alpha} & z^{2-{2}\alpha}+z^{-2}
         \end{array}\right)  \text{with prob.} \ 1-x \vspace{0.1 cm} \\
               \left(\begin{array}{cc}
            z^{-1+\alpha} & z^{1+\alpha} \\
            z^{1-\alpha} & z^{-1-\alpha}
          \end{array}\right)  \text{with prob.} \ x\
             \end{array}  \right.  \label{ffa} \\ \nonumber \\
 \text{AAF:} \hspace{0.5 cm} \mathcal{L}_i &=& \left\{\begin{array}{l}
                \left(\begin{array}{cc}
            z^{1+\alpha} & z^{-1+\alpha} \\
            z^{-1-\alpha} & z^{1-\alpha}
         \end{array}\right) \text{with prob.} \ 1-x  \vspace{0.1 cm} \\
                \left(\begin{array}{cc}
            z^{-2+2\alpha}+z^{2} & z^{2\alpha}+1 \\
            1+z^{-2\alpha} & z^{-2-2\alpha}+z^{2}
          \end{array}\right)  \text{with prob.} \ x
             \end{array}  \right. \label{faa}
\end{eqnarray}
It should be noted that the average quantities that we introduced above will not be per spin if we combine pairs in a single matrix and we will need the factor $\frac{N_{\text{matrix}}}{N_{\text{spin}}}$ to transform per matrix quantities to per spin. According to the above distribution of matrices, this ratio is $\frac{1}{2-x}$ and $\frac{1}{1+x}$ for AFF and AAF models respectively.

The above distribution of transfer matrices introduces exponentially decaying correlations between couplings with the following mean and correlation functions
\begin{eqnarray}
&&\text{AFF:}   \left\{\begin{array}{l} \langle J_i \rangle =  J(2-3x)/(2-x),\label{Jmean1}\\
   \langle J_i J_{i+n}\rangle - \langle J_i \rangle^2 = 4J^2 \textstyle{\left(\frac{ x}{2-x}\right)^2} (-1)^{n+1} (1-x)^n ,\end{array}\right.\label{Jcorr1}
  \vspace{0.5 cm}\\
&&\text{AAF:}    \left\{\begin{array}{l} \langle J_i \rangle =  J(1-3x)/(1+x),\label{Jmean2}\\
   \langle J_i J_{i+n}\rangle - \langle J_i \rangle^2 = 4J^2 \textstyle{\left(\frac{ 1-x}{1+x}\right)^2} (-1)^{n+1} x^n .\end{array}\right.\label{Jcorr2}
\end{eqnarray}
The respective correlation lengths of couplings in two models are $\log \frac{1}{1-x}$ and $\log \frac{1}{x}$ which can be adjusted by changing the concentration of dimers.
\section{Stochastic recursive equations and stationary distribution functions}

In this section, we obtain the recursive equations for exponents and determine the values that $c$ and $w$ take by iteration of these equations. Then the question is how often we encounter each of these values, i.e., what is the limit probability of occurrence of them. Using the fact that the stationary probability density must be invariant under recursive equations, we obtain a set of equations for probabilities.

\subsection{\textbf{AFF model}}
Using Eqs. (\ref{rec-eq}) and (\ref{ffa}) we have with probability $1-x$:
\begin{eqnarray}
\text{If} \ -2\leq c<2-4\alpha:&& \left\{ \begin{array}{lll}
            a'=a+2+2\alpha   & \ \ u'=u \\
            b'=b+2-2\alpha   & \ \ v'=v\\
            c'=c+4\alpha   & \ \ w'=w.
         \end{array}\right.\\
\text{If} \ c=2-4\alpha  :&&  \left\{\begin{array}{lll}
            a'=a+2+2\alpha   & \ \ u'=u \\
            b'=b+2-2\alpha   & \ \  v'=\ln(e^{u}+e^{v})\\
            c'=2   & \ \  w'=-\ln(1+e^{-w}).
         \end{array}\right.\\
\text{If} \ 2-4\alpha<c\leq2  &:&  \left\{\begin{array}{lll}
            a'=a+2+2\alpha   & \ \  u'=u \\
            b'=b+2-2\alpha   & \ \  v'=u\\
            c'=2   & \ \  w'=0.
         \end{array}\right.
\end{eqnarray}
and with probability $x$:
\begin{eqnarray}
\text{If} \ -2\leq c<2-2\alpha  &:& \left\{ \begin{array}{lll}
            a'=b+1-\alpha   &&  u'=v \\
            b'=a+1+\alpha   &&  v'=u\\
            c'=-c-2\alpha   &&  w'=-w.
         \end{array}\right.\\
\text{If} \ c=2-2\alpha  &:&  \left\{ \begin{array}{lll}
            a'=b+1-\alpha   &&  u'=\ln(e^{u}+e^{v}) \\
            b'=a+1+\alpha   &&  v'=u \\
            c'=-2   &&  w'=-\ln(1+e^{-w}).
         \end{array}\right.\\
\text{If} \ 2-2\alpha<c\leq2  &:& \left\{ \begin{array}{lll}
            a'=a-1+\alpha   &&  u'=u \\
            b'=a+1+\alpha   &&  v'=u\\
            c'=-2  &&  w'=0.
         \end{array}\right.
\end{eqnarray}
As we can see, $c$ takes the values $\pm 2$, can only change by amount $2\alpha$, and according to the preconditions, is restricted to the interval $-2\leq c\leq2$. Therefore, the possible values of $c$ are $-2,-2+2\alpha,\cdots,2-2\alpha,2$. Let $p_i$ be the probability of having $c=-2+2i\alpha$ and $q_i$ be the probability that $c=2-2i\alpha$ where $i$ is a non-negative integer whose maximum value depends on $\alpha$. For $\frac{2}{r+1}<\alpha<\frac{2}{r}$, we have $0\leq i \leq r$ where $r \geq 0$ is an integer. The invariant probabilities under the above recursive equations must satisfy
\begin{eqnarray}
\begin{array}{ll}
  p_{0}=x(p_{r}+q_{1}+q_{0}) & q_{0}=\frac{1-x}{x}(p_{r}+p_{r-1}+q_{1}+q_{2}) \\
  p_{1}=x q_{2} & q_{r-1}=x p_{r-2} \\
  p_{r}=(1-x)p_{r-2} & q_{r}=x p_{r-1} \\
  p_{i}=(1-x)p_{i-2}+x q_{i+1},\  2\leq i\leq r-1, \ \ \ \ & q_{i}=(1-x)q_{i+2}+x p_{i-1},\  1\leq i\leq r-2,
\end{array}
\end{eqnarray}
and an additional equation to ensure the normalization $\sum_{i}(p_{i}+q_{i})=1$. This set of equations allows the following closed solution
\begin{eqnarray}
r \ \text{even} \left\{
\begin{array}{l}
  p_i=\frac{x[2+(r-i-2)x][1+(-1)^i]}{[2+(r-2)x](2+rx)}, \ 0\leq i \leq r \\
  q_i=\frac{x^2(r-i+1)[1+(-1)^{i-1}]}{[2+(r-2)x](2+rx)}, \ 1\leq i\leq r \\
  q_0=\frac{2(1-x)}{2+rx},
\end{array}
\right.
\ \ \ r \ \text{odd} \left\{
\begin{array}{l}
  p_i=\frac{x[2+(r-1-i)x][1+(-1)^i]}{[2+(r-1)x][2+(r+1)x]}, \ 0\leq i \leq r \\
  q_i=\frac{x^2(r-i+2)[1+(-1)^{i-1}]}{[2+(r-1)x][2+(r+1)x]}, \ 1\leq i\leq r \\
  q_0=\frac{2(1-x)}{2+(r-1)x}.
\end{array}
\right.
\end{eqnarray}
\subsection{\textbf{AAF model}}
For this case, we have with probability $1-x$:
\begin{eqnarray}
\text{If} \ -2\leq c<2-2\alpha:&&\
\left\{\begin{array}{ll}
a'=a+1+\alpha & \ \ u'=u \\
b'=b+1-\alpha & \ \ v'=v \\
c'=c+2\alpha  & \ \ w'=w.
\end{array}
\right. \\
\text{If} \ c=2-2\alpha:&&\
\left\{\begin{array}{ll}
a'=a+1+\alpha & \ \ u'=u \\
b'=b+1-\alpha & \ \ v'=\ln(e^{u}+e^{v}) \\
c'=c+2\alpha  & \ \ w'=-\ln(1+e^{-w}).
\end{array}
\right.\\
\text{If} \ 2-2\alpha<c\leq2:&&\
\left\{\begin{array}{ll}
a'=a+1+\alpha & \ \ u'=u \\
b'=a-1+\alpha & \ \ v'=u \\
c'=2  & \ \ w'=0.
\end{array}
\right.
\end{eqnarray}
and with probability $x$:
\begin{eqnarray}
\text{If} \ c=-2:&&\
\left\{\begin{array}{ll}
a'=a+2 & \ \ u'=\ln(e^{u}+e^{v}) \\
b'=b+2 & \ \ v'=v \\
c'=-2  & \ \ w'=\ln(1+e^{-w}).
\end{array}
\right. \\
\text{If} \ -2<c<2-2\alpha:&&\
\left\{\begin{array}{ll}
a'=a+2 & \ \ u'=u \\
b'=b+2 & \ \ v'=v \\
c'=c  & \ \ w'=w.
\end{array}
\right.\\
\text{If} \ c=2-2\alpha:&&\
\left\{\begin{array}{ll}
a'=a+2 & \ \ u'=u \\
b'=b+2 & \ \ v'=\ln(e^{u}+e^{v}) \\
c'=2-2\alpha  & \ \ w'=-\ln(1+e^{-w}).
\end{array}
\right.\\
\text{If} \ 2-2\alpha<c\leq2:&&\
\left\{\begin{array}{ll}
a'=a+2 & \ \ u'=u \\
b'=a+2\alpha & \ \ v'=v \\
c'=2-2\alpha  & \ \ w'=0.
\end{array}
\right.
\end{eqnarray}
In this case, we have much simpler invariant probability equations,
\begin{eqnarray}
            p_{i}&=&0 , \ \ \  0\leq i\leq r,\\
            q_{i}&=&0 , \ \ \ 2\leq i\leq r, \\
            q_{0}&=&\frac{x}{1-x}q_{1}.
\end{eqnarray}
Using the normalization condition, $\sum_{i}(p_{i}+q_{i})=1$, we have $q_{0}=x$, $q_{1}=1-x$ and zero for all other probabilities.
\section{Ground state Energy and Magnetization}

Using the recursive equations and identities (\ref{energy}) and  (\ref{ava}), we can now obtain the ground state energy, magnetization, and susceptibility. For AFF model, the ground state energy is given by
\begin{eqnarray}
E_{\text{AFF}}/J=\langle \Delta a \rangle &=& \frac{1-x}{2-x}\sum_{i=0}^{r}(p_i+q_i)(2+2\alpha)+\frac{x}{2-x}\sum_{i=0}^{r-1}p_{i}(2-2i\alpha+1-\alpha) \nonumber\\
&& \hspace{1.5cm} +\frac{x}{2-x}\sum_{i=1}^{r}q_{i}(-2+2i\alpha+1-\alpha)+\frac{x}{2-x}(p_{r}+q_{0})(\alpha-1),\nonumber\\
&=&\left\{\begin{array}{l}
                \frac{4 (1 - x) [2 + (2 r-1) x]\alpha + x [  r (2 - x) (4 + (r-2) x) +12 x - 20] +8 }{(2-x)(2+r x)[2+(r-2)x]} \ \text{even} \ r
                \vspace{0.2 cm}\\
                \frac{4 (1 - x) [2 + (2 r+1) x]\alpha + x [  r (2 - x)(4 + r x)+ x (6 + x)- 12] +8 }{(2-x)[2+(r-1)x][2+(r+1)x]} \ \text{odd} \ r\
             \end{array}  \right.,
\end{eqnarray}
and using the energy, magnetization and susceptibility are obtained as
\begin{eqnarray}
&&M_{\text{AFF}}=\frac{\partial E_{\text{AFF}}}{\partial h}=\left\{\begin{array}{l}
                \frac{4(1-x)[2+(2r-1)x]}{(2-x)(2+r x)[2+(r-2)x]} \ \text{even} \ r
                \vspace{0.2 cm}\\
                \frac{4(1-x)[2+(2r+1)x]}{(2-x)[2+(r-1)x][2+(r+1)x]} \ \text{odd} \ r\
             \end{array}  \right., \\
&&\chi_{\text{AFF}}(0)=\lim_{h\rightarrow 0} \frac{M_{AFF}}{h} =\frac{4(1-x)}{J x(2-x) }.
\end{eqnarray}
Similarly for AAF model we have
\begin{eqnarray}
&&E_{\text{AAF}}/J=\frac{1}{1+x}[(1+\alpha)(1-x)+2x)], \\
&&M_{\text{AAF}}=\frac{\partial E_{\text{AAF}}}{\partial h}=\frac{1-x}{1+x}.
\end{eqnarray}

For all three models, AF, AFF and AAF when $\alpha>2$, $c$ only takes the values $\pm2$ and the energy per spin is given by $E=\langle J \rangle+h$, therefore, the magnetization per spin is $M=1$ which is clear from the fact that all spins align fully along the field direction.
\begin{figure}
\center \epsfxsize8truecm \epsffile{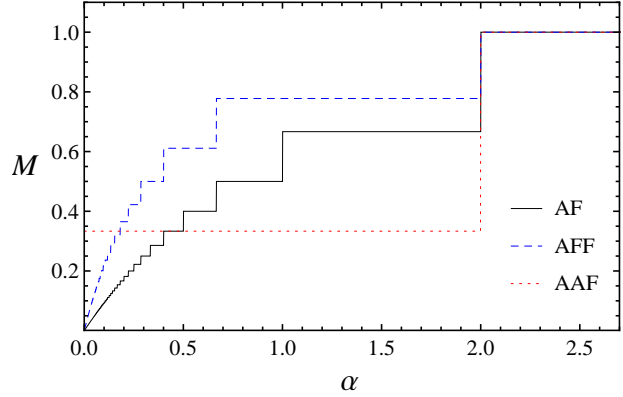}
\caption{Ground State magnetization of the models AF (solid), AFF (dashed) and AAF (dotted) for $x=0.5$.}
\label{figu1}
\end{figure}

Figure \ref{figu1} shows the magnetization of AFF and AAF models compared with AF model as a function of magnetic field. Magnetization steps at critical fields with even denominator, $\alpha=\frac{2}{2n}$, $n=1,2,\cdots$, are absent in the AFF model. This can be understood in terms of clusters of down spins. In the case of the AFF model, since we only have dimers of ferromagnetic bonds, there are only an even number of ferromagnetic couplings in succession, which result in clusters with an odd number of spins. The AAF model does not show any jump in magnetization except at $\alpha=2$. Again, this is because the antiferromagnetic bonds come in pairs and the clusters of ferromagnetic sections on two sides of an antiferromagnetic section should align in the same direction. This is because in passing from one cluster to another the spins must flip an even number of times. Thus, the only clusters of down spins are single spins which have turned over due to antiferromagnetic couplings and all of them flip when magnetic field exceeds $h=2J$.

\begin{acknowledgements}
We would like to thank Bernard Derrida for illuminating discussion and pointing out the possibility of flipping a combination of clusters at the critical fields. We also thank Ehsan Khatami for critical reading of the manuscript.
\end{acknowledgements}

\bibliographystyle{unsrt}
\bibliography{refs}   % name your BibTeX data base

\end{document}